\documentclass[aps,groupaddress,twocolumn,letterpaper]{revtex4}
\usepackage{graphicx}
\usepackage{amsmath}
\usepackage{dcolumn}
\usepackage{subfigure}
\usepackage{multirow}
\usepackage{mathtools}
\usepackage{mathptmx}
\usepackage{braket}
\begin{document}
	
\title{Optomechanics and thermometry of cryogenic silica microresonators}
\author{A. J. R. MacDonald, B. D. Hauer, X. Rojas, P. H. Kim, G. G. Popowich and J. P. Davis}
\date{\today}
\affiliation{Department of Physics, University of Alberta, Edmonton, AB, Canada T6G 2G7}
	
\begin{abstract}	
We present measurements of silica optomechanical resonators, known as bottle resonators, passively cooled in a cryogenic environment. These devices possess a suite of properties that make them advantageous for preparation and measurement in the mechanical ground state, including high mechanical frequency, high optical and mechanical quality factors, and optomechanical sideband resolution. Performing thermometry of the mechanical motion, we find that the optical and mechanical modes demonstrate quantitatively similar laser-induced heating, limiting the lowest average phonon occupation observed to just $\sim$1500. Thermalization to the 9 mK thermal bath would facilitate quantum measurements on these promising nanogram-scale mechanical resonators. 
\end{abstract}

\maketitle
	
\section{Introduction}

Cavity optomechanical systems, consisting of high-quality optical cavities coupled to mechanical resonators, offer a promising route to making precision measurements in both the applied and fundamental science domains \cite{Aspelmeyer2014}. In dispersively-coupled systems, the motion of the mechanical resonator shifts the resonance frequency of the optical cavity, which can in turn be observed as periodic changes in the amplitude or phase of light traversing the cavity. Optomechanics thus provides an extremely sensitive readout for micro- and nano-mechanical resonators, enabling their use as exquisite sensors of a variety of phenomena on small scales, including displacement \cite{Anetsberger2010}, force \cite{Gavartin2012,Miao2012,Doolin2014}, torque \cite{Kim2013,Wu2014} and acceleration \cite{Krause2012}. This readout sensitivity has also motivated fundamental searches for quantum properties of nanomechanical resonators \cite{Gangat2010} at, or near, their vibrational ground state where mechanical motion originates from quantum zero-point fluctuations. The hybrid nature of optomechanical systems additionally makes them desirable for applications in quantum information processing architectures \cite{Stannigel2010}, and as a resource for entangled \cite{Palomaki2013}, squeezed \cite{SafaviNaeini2013} or other non-classical states \cite{Borkje2014}.

Noise from thermally-driven oscillations of mechanical resonators poses a significant obstacle to performing many of these proposed experiments. Mechanical resonators with frequencies in the kHz to GHz range have thermal occupancies of billions to thousands of phonons at room temperature, which can easily drown out any quantum signature. Reducing this thermal occupation is thus of great importance. The optomechanical interaction can be exploited to actively cool the mechanical mode into or near its ground state \cite{Schliesser2008,Chan2011,Park2009} but a fundamental limit to this process is imposed by the temperature of the mechanical resonator's bath, necessitating cryogenic pre-cooling \cite{Aspelmeyer2014}. Active optomechanical cooling furthermore reduces the quality factor ($Q_\text{m}$) of the mechanical resonator, decreasing the signal-to-noise ratio at low temperatures \cite{Aspelmeyer2014}. For these reasons, we focus on passively cooling our resonators, which additionally facilitates the use of sensitive optomechanical systems for probing low temperature phenomena such as superfluidity \cite{Sun2013,DeLorenzo2014} and superconductivity \cite{Geim1997}.

\begin{figure}[b]
	\includegraphics[width=3.3 in]{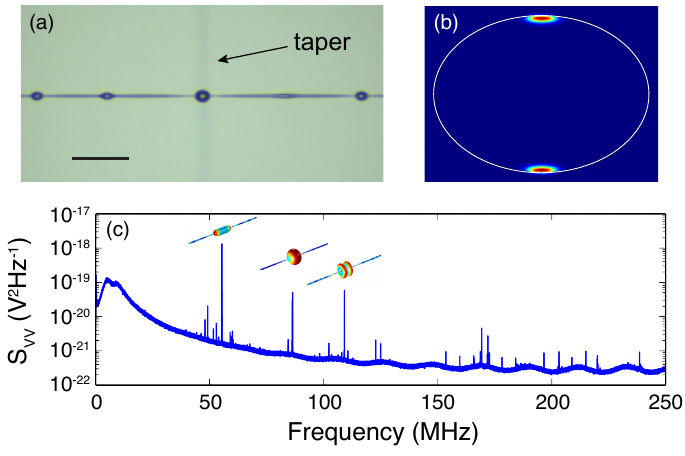}
	\caption{(a) Photograph of a string of bottle resonators along with the tapered fiber for optomechanical measurement. Scale bar is 100 $\mu$m. (b) Simulation of a whispering gallery mode optical resonance, showing a cross-section of the bottle with the optical field localized along the equator. (c) Representative mechanical spectrum, measured at room temperature. The three most prominent resonances are indicated by the corresponding mechanical mode simulation.}
	\label{fig:fig1}
\end{figure}	

In previous experiments, optomechanical systems have been passively cooled using both helium flow cryostats \cite{Park2009,Riviere2013} and dilution refrigerators \cite{Palomaki2013,Meenehan2014}. Our system uses a dilution refrigerator for its lower achievable base temperature and, in contrast to the system presented in Ref.~\citenum{Meenehan2014}, uses the highly-efficient tapered-fiber coupling method \cite{Knight1997,Cai2000}, which is compatible with both bulk and on-chip optomechanical systems \cite{MacDonald2015,Hauer2014}. 

Here, we present measurements of so-called silica bottle resonators \cite{Kakarantzas2001,Pollinger2009}, shown in Figure \ref{fig:fig1}, which have ellipsoidal shapes and exhibit optical whispering gallery modes (WGM) in the infrared and visible range, a simulation of which is shown in Figure \ref{fig:fig1}(b). They are interesting structures for their high optical quality factors ($Q_\text{o}$ up to $10^{8}$) and the tunability of their mode structure \cite{Pollinger2009}. Additionally, these bottle resonators have mechanical breathing modes with frequencies in the 50 to 250 MHz range, Figure \ref{fig:fig1}(c), corresponding to average phonon occupancies of just 3.3 and 0.4, respectively, at a temperature of 9 mK (the base temperature of our dilution refrigerator). The significant modal overlap of the optical and mechanical modes, which are both localized at the equator, leads to a large optomechanical coupling, on the order of $G/2\pi\sim10$ GHz/nm. Furthermore, the combination of high optical quality factor and high mechanical frequency places these devices in the sideband-resolved regime, enabling many of the important tools of optomechanics \cite{Aspelmeyer2014}.

\section{Experimental Details}

We fabricate our bottle resonators by applying tension to each end of a single mode optical fiber while melting it with a CO$_2$ laser. This process results in a string of resonators, as shown in Figure \ref{fig:fig1}(a), separated from each other by thin stems. By controlling the intensity of the CO$_2$ laser and the length of the pull time, we create resonators with symmetric shapes and extremely thin supporting stems. This reduces phonon tunnelling through the stems and increases the mechanical quality factor \cite{Anetsberger2008}.

Tapered fibers \cite{Knight1997,Cai2000} are used to couple light into the WGMs of the bottles. These fibers have a small core (on the order of the wavelength of the light) and no cladding, save the medium surrounding the fiber. Our tapers are created by heating an optical fiber with a hydrogen torch and pulling it until a minimum diameter of $\sim$1 $\mu$m is reached \cite{Hauer2014}, resulting in a large evanescent field. This allows efficient coupling of light to WGM resonators through frustrated total internal reflection.

\begin{figure}
	\includegraphics[width=\columnwidth]{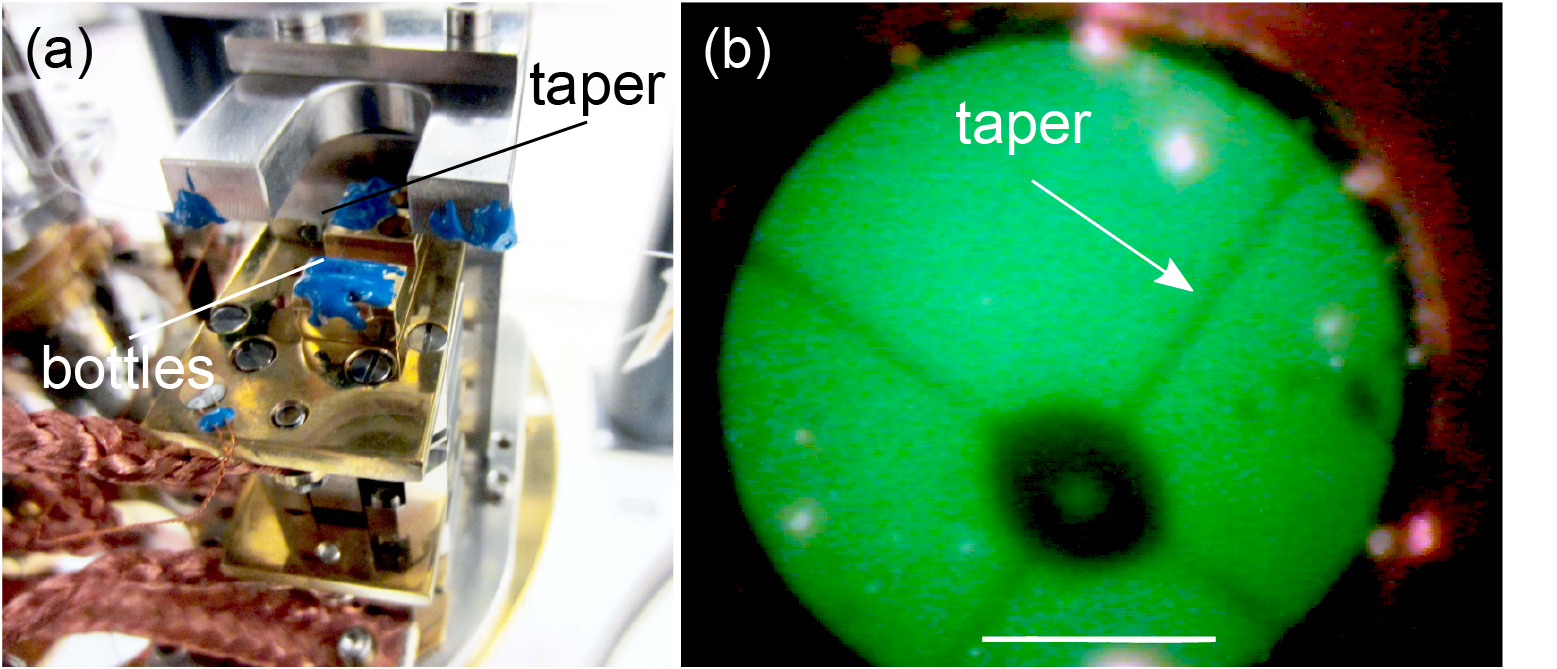}
	\caption{(a) Optomechanical coupling apparatus on the mixing chamber stage of our dilution refrigerator. (b) Image of a bottle resonator and tapered fiber taken with the low-temperature microscope. Scale bar is 100 $\mu$m.}
	\label{fig:bottles}
\end{figure}

We passively cool the taper and bottle resonators using a dilution refrigerator with a base temperature of 9 mK \cite{MacDonald2015}. To facilitate taper-resonator coupling at cryogenic temperatures, we have built an optomechanical coupling system within the inner vacuum can (IVC), on the mixing chamber plate, as pictured in Figure \ref{fig:bottles}(a). The bottles are mounted on top of a stack of nanopositioning stages, which allow full three-dimensional control over the position of the resonators. They are thermally anchored to the mixing chamber through a series of oxygen-free high-conductivity copper plates and braids. The taper is epoxied and mounted to a large Invar block, which minimizes the effects of thermal contractions. A homebuilt low-temperature microscope \cite{MacDonald2015} enables real-time \emph{in situ} imaging of the taper-resonator system with a resolution of $\sim$1 $\mu$m (Figure \ref{fig:bottles}(b)). We inject light from a tunable diode laser housed at room temperature into the tapered fiber and detect the transmitted light with a fast photodetector. The optical and mechanical properties of the bottles can thus be measured from the low- and high-frequency parts of the photodetector voltage, respectively.

\section{Optical Properties}

Characterization of the near-infrared optical modes is performed by scanning a tunable diode laser (1500-1630 nm) with angular frequency $\omega_{\text{L}}$ across a resonance, resulting in the following expression for the low-frequency part of the transmission $\dot{n}_{\text{out}}$ (in units of photons per second) through the tapered fiber

\begin{equation}
\dot{n}_{\text{out}}= \dot{n}_\text{in}-\kappa_0n_{\text{cav}}.
\label{eq:nout}
\end{equation}
	
\noindent Here $\dot{n}_\text{in}=P_{\text{in}}/\hbar\omega_{\text{L}}$ is the rate of photons passing through the tapered fiber, for injected laser power $P_{\text{in}}$ and reduced Planck constant $\hbar$. The linewidth of the resonance is given by $\kappa=\kappa_0+\kappa_\text{ex}$ where $\kappa_0$ is the intrinsic decay rate and $\kappa_\text{ex}$ is the rate at which photons are exchanged between the bottle and tapered fiber. Our setup allows us to sensitively control the relative strengths of these intrinsic and extrinsic decay rates \cite{MacDonald2015}. For the measurements that follow, the resonator is operated in the slightly-undercoupled regime ($\kappa_\text{ex}\lesssim\kappa_0$).

The number of photons inside the cavity, 

\begin{equation}
n_{\text{cav}}=\frac{\dot{n}_\text{in}\kappa_{\text{ex}}}{\frac{\kappa^2}{4}+\Delta^2},
\label{eq:n_cav}
\end{equation}

\noindent is a function of the laser detuning $\Delta=\omega_{\text{L}}-\omega_{\text{o}}$ from the cavity resonance, $\omega_\text{o}$. For the whispering gallery modes in a bottle of radius $R$, the angular optical resonance frequencies $\omega_{\text{o}}$ are given by

\begin{equation}
\omega_{\text{o}}=\frac{lc}{Rc_1n}.
\label{eq:resonance_cond}
\end{equation}

\noindent The integer $l$ is a mode label for a particular resonance, $c$ is the speed of light in vacuum, $n$ is the refractive index of the bottle and $c_1$ is a geometric factor which accounts for the fact that the optical mode is not perfectly confined to the surface of the resonator \cite{Carmon2004}. 

For relatively low injected optical powers ($P_{\text{in}}\le250$ nW), the optical resonances can be measured without heating the mixing chamber thermometer and have $Q_\text{o}$ of 10$^6$-10$^7$, which are consistent with their room-temperature quality factors. At larger injected powers, light lost from the taper ($\sim$30\%) heats the cryogenic environment \cite{MacDonald2015}. Furthermore, the high $Q_\text{o}$, and correspondingly long photon lifetime, in the cavity leads to an increased absorption of laser photons in the silica. 

\begin{figure}
	\includegraphics[width=\columnwidth]{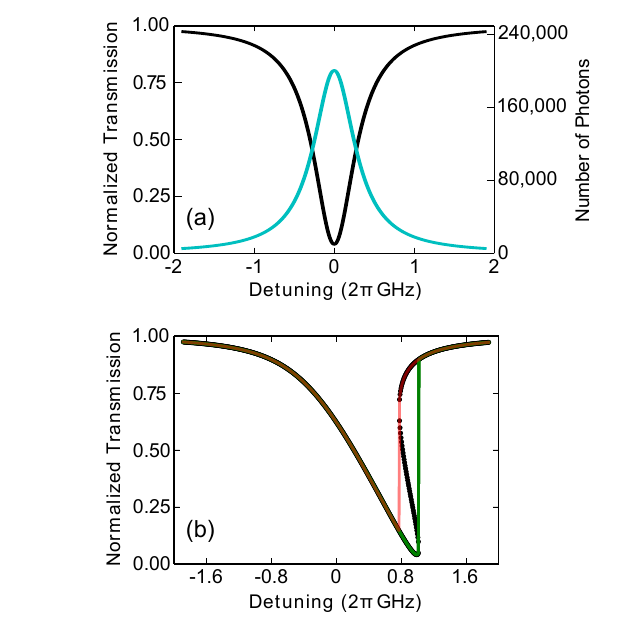}
	\caption{(a) Simulated normalized taper transmission (black) and number of intracavity photons (blue) for a linear optical resonance with parameters comparable with those seen in experiment ($\kappa/2\pi=50$ MHz, $\kappa_\text{ex}/2\pi=20$ MHz and $P_\text{in}=4$ $\mu$W). (b) Simulation of a nonlinear resonance, with the same parameters as above and $\Delta_\text{nl}/2\pi=-5\times10^{-6}$ Hz. All real solutions to Equation \eqref{eq:n_cav_nonlin} are plotted in black, while the transmission for a forward (reverse) scan of the laser is plotted in green (red).}
	\label{fig:nonlin_theory}
\end{figure}

This absorption generates a local heating of the silica within the optical mode volume, which in turn causes both a change in the bottle dimensions through thermal expansion and a change in the refractive index through the thermo-refractive effect. We can then rewrite Equation \eqref{eq:resonance_cond} with Taylor-expanded expressions for the radius and the refractive index \cite{Carmon2004},

\begin{equation}
\begin{split}
\omega_{\text{o}}(T) &= \frac{lc}{R(1+\epsilon\Delta T)c_1n(1+\frac{1}{n}\frac{\partial n}{\partial T}\Delta T)} \\
&\approx\omega_{\text{o}}(1-b(T)\Delta T)
\label{eq:omega_T}
\end{split}
\end{equation}

\noindent where $T$ is the equilibrium temperature of the glass in the absence of laser heating, $\Delta T$ is the temperature increase caused by optical absorption, and $\epsilon$ is the linear thermal expansion coefficient. Although higher-order terms become important under some conditions \cite{Arcizet2009}, here we keep only terms to first order in $\Delta T$ and for simplicity write $b(T) = \epsilon+\frac{1}{n}\frac{\partial n}{\partial T}$.

Incorporating the modified $\omega_{\text{o}}$ into Equation \eqref{eq:n_cav}, we have

\begin{equation}
n_{\text{cav}}=\frac{\dot{n}_\text{in}\kappa_{\text{ex}}}{\frac{\kappa^2}{4}+(\Delta+\Delta_{\text{nl}}n_{\text{cav}})^2},
\label{eq:n_cav_nonlin}
\end{equation}

\noindent where the thermo-optical effects have been written into a nonlinear detuning parameter $\Delta_\text{nl}$ \cite{Barclay2005,Doolin2014b}. This parameter can be thought of as the shift in the optical resonance frequency per intracavity photon, and is related to the parameters in Equation \eqref{eq:omega_T} by $\Delta_{\text{nl}}n_{\text{cav}}=b(T)\Delta T$, under the assumption that the temperature gradient $\Delta T$ is proportional to the number of intracavity photons. As we will see, the exact constant of proportionality will depend on the resonator's heat dissipation.

Equation \eqref{eq:n_cav_nonlin} is now a nonlinear function of $n_{\text{cav}}$, giving rise to up to three distinct real solutions for appropriate values of $\Delta_{\text{nl}}$. Experimentally, this is manifest as a bistability and hysteresis in the resonance shape, which is dependent on the scanning direction of the laser \cite{Braginsky1989,Illchenko1992}, as illustrated in Figure \ref{fig:nonlin_theory}. As the laser is tuned closer to the optical resonance, more photons are coupled into the cavity, causing a greater shift in the resonance frequency, such that it becomes distorted from the linear shape shown in Figure \ref{fig:nonlin_theory}(a).

\begin{figure}[t]
	\includegraphics[width=\columnwidth]{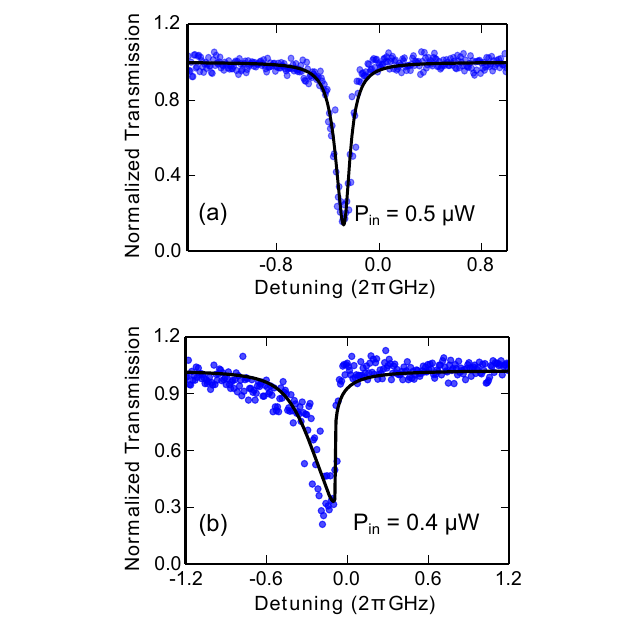}
	\caption{Transmission measured across a bottle optical resonance with a wavelength of 1516 nm and a temperature of 4.2 K. For similar injected powers, the resonance exhibits different behavior when the bottle is (a) immersed in helium exchange gas and (b) surrounded by vacuum. Fits to the data (black line) yield nonlinear detuning parameters $\Delta_\text{nl}/2\pi$ of approximately (a) $-3\times10^{-7}$ Hz and (b) $-2\times10^{-6}$ Hz.}
	\label{fig:nonlin_exp}
\end{figure}

This nonlinear optical behaviour is observed in our experiments, as shown in Figure \ref{fig:nonlin_exp} for a $R=25$ $\mu$m bottle. Using the cubic equation to exactly solve Equation \eqref{eq:n_cav_nonlin}, we fit our experimental data to extract $\Delta_{\text{nl}}$ under various experimental conditions. In particular, we find that injecting a large quantity of helium exchange gas into the IVC at 4.2 K reduces $\Delta_{\text{nl}}$ by an order of magnitude ($\Delta_{\text{nl}}/2\pi\approx-3\times10^{-7}$ Hz compared to $-2\times10^{-6}$ Hz in vacuum). We furthermore find that $\Delta_{\text{nl}}$ at the fridge base temperature ($\Delta_{\text{nl}}/2\pi\approx-3\times10^{-6}$ Hz) is comparable to that at 4.2 K in vacuum. 

Although it will not be discussed here, it is worth noting that at intermediate pressures of helium exchange gas, we observe a higher-order thermo-optical nonlinearity. This results in up to five real solutions to a modified form of Equation \eqref{eq:n_cav_nonlin}, and presents itself in the experiment as a multistability in the resonance profile. This effect is thought to originate from a reversal of the thermo-refractive effect at low temperatures \cite{Arcizet2009}, although the nonlinear temperature dependence of the thermal expansion coefficient \cite{White1975} may also contribute.

\section{Mechanical Properties}

The high-frequency part of the taper transmission encodes the mechanical motion of the bottle resonator. We collect this signal by recording a time trace of the high-pass filtered photodetector voltage with a fast analog-to-digital converter. The $R=25$ $\mu$m bottle studied has several mechanical resonances, with the most prominent being at 55, 85 and 109 MHz, with room temperature quality factors of $\sim$ $10^4$. At low temperatures, we focus on the lowest-frequency (55 MHz) mode, which has an effective mass $m_\text{eff}=64$ ng and optomechanical coupling strength $G/2\pi\sim8$ GHz/nm.

It is important to note that, despite significant efforts to thermally anchor the bottle resonators to the base plate of the fridge, the thermometers used to measure the temperature of the mixing chamber will not provide an accurate measure of the mechanical mode temperature. The thermally-insulating nature of silica along with the very thin connections between the bottles prevent efficient conduction of heat, while incident laser light used to measure the mechanics leads to local heating of the resonator. Dynamical back-action effects in the optomechanical interaction can also lead to mode-specific heating or cooling \cite{Aspelmeyer2014} which is not reflected in the temperature of the bulk silica. 

To independently determine the temperature, we exploit the thermally-driven motion of the resonator. In this case, the resonator's spectral response is quantified by its one-sided power spectral density (PSD). For a resonator with effective mass, $m_\text{eff}$, at temperature $T$, the displacement PSD is given by \cite{Hauer2013}

\begin{equation}
\begin{split}
S_{xx}(\omega)&=\lim\limits_{\tau\to\infty}\frac{1}{\tau}|X(\omega)|^2 \\
&=\frac{4k_{\text{B}}T\Omega_{\text{m}}}{m_{\text{eff}}Q_{\text{m}}\left[(\Omega_{\text{m}}^2-\omega^2)^2+(\omega\Omega_{\text{m}}/Q_{\text{m}})^2\right]},
\end{split}
\end{equation}

\noindent for a sufficiently long measurement time $\tau$ ($\tau\gg2\pi/\Omega_\text{m}$). Here, $k_\text{B}$ is the Boltzmann constant, $\Omega_\text{m}$ is the angular frequency of the mechanical resonance and $X(\omega)$ is the Fourier transform of the resonator's position $x(t)$. It is evident that the amplitude of the displacement PSD ($4k_\text{B}T/m_\text{eff}$) scales linearly with $T$ and thus gives a direct measurement of the mode temperature. 

In our experiments, we measure this displacement response by Fourier transforming the photodetector voltage, squaring and dividing the result by the measurement bandwidth. The voltage PSD is given by the sum of a detection system-dependent noise floor $S_{VV}^\text{NF}$ and the transduction of $S_{xx}(\omega)$ through the optical cavity and photodetector,

\begin{equation}
S_{VV}(\omega)=S_{VV}^{\text{NF}}+\dot{n}_\text{in}^2\frac{G^2}{\Omega_{\text{m}}^2}\alpha(\Omega_{\text{m}},\Delta)S_{xx}(\omega).
\label{eq:Svv}
\end{equation}

\noindent The transduction of $S_{xx}(\omega)$ scales with the square of the rate of injected photons, as well as the square of the ratio of the optomechanical coupling strength to the mechanical resonance frequency. All of the detuning dependence of $S_{VV}(\omega)$ is contained in $\alpha(\Omega_\text{m},\Delta)$ (see Appendix), a function which describes the transduction of the resonator's fluctuations in position into fluctuations in first taper transmission and then voltage.

\begin{figure}
	\includegraphics[width=\columnwidth]{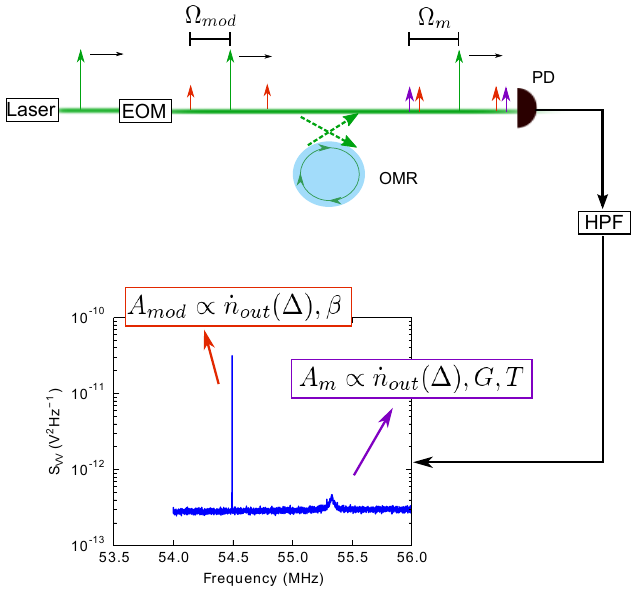}
	\caption{Schematic of the mechanical mode thermometry technique. Modulation of laser light by an electro-optic modulator (EOM) generates optical sidebands on the laser spaced by $\Omega_\text{mod}$. The motion of the optomechanical resonator (OMR) imposes additional sidebands, at $\Omega_\text{m}$. Both sets of sidebands are detected by the photodiode (PD), resulting in a high-pass filtered (HPF) spectrum that has two distinct peaks corresponding to the phase modulation (54.5 MHz) and the mechanical motion (55 MHz).}
	\label{fig:phasecalibration}
\end{figure}

Precise knowledge of the laser detuning, along with the optomechanical coupling strength $G$, injected optical power and the gains of the detection electronics, would allow the extraction of $S_{xx}(\omega)$, and hence the temperature, from Equation \eqref{eq:Svv}. This is not always experimentally feasible, so we instead indirectly calibrate $S_{VV}(\omega)$ by injecting an all-optical signal of known energy into the system \cite{Gorodetsky2010}. The general procedure for doing so is outlined in Figure \ref{fig:phasecalibration}. An electro-optic modulator (EOM) driven at angular frequency $\Omega_\text{mod}$ phase modulates the input laser light, generating optical sidebands on the laser at frequencies $\omega_\text{L}\pm k\Omega_\text{mod}$ for integer $k$. If the phase modulation depth, $\beta$, is small ($\beta=\pi V_0/V_\pi\ll1$ for applied voltage $V_0$ and EOM half-wave voltage $V_\pi$), only the first order sidebands ($k=1$) need be considered. If $\Omega_\text{mod}$ is chosen such that $|\Omega_\text{mod}-\Omega_\text{m}|\gg\Gamma$ where $\Gamma=\Omega_\text{m}/Q_\text{m}$ is the resonance linewidth, the PSD of the photodetector signal becomes (see Appendix)

\begin{equation}
\begin{split}
S_{VV}(\omega)=& S_{VV}^{\text{NF}}+\frac{P_{\text{in}}^2}{\hbar^2\omega_{\text{L}}^2}\left(\frac{G^2}{\Omega_{\text{m}}^2}\alpha(\Omega_{\text{m}},\Delta)S_{xx}(\omega)\right.\\
&+\alpha(\Omega_{\text{mod}},\Delta)S_{\phi\phi}(\omega)\Big),
\end{split}
\end{equation}

\noindent where $S_{\phi\phi}(\omega)$ is the spectrum of the applied phase modulation. This spectrum features two distinct peaks, as shown in Figure \ref{fig:phasecalibration} at $\Omega_\text{mod}$ and $\Omega_\text{m}$.

If we additionally assume that $\Omega_\text{mod}\approx\Omega_\text{m}$, such that the response of the detection electronics is flat, then $\alpha(\Omega_\text{mod},\Delta)\approx\alpha(\Omega_\text{m},\Delta)$ and the phase modulation signal and mechanical motion are transduced similarly through the optical cavity. Although the strength of each peak in $S_{VV}(\omega)$ depends sensitively on the injected optical power and the laser detuning, this dependence is identical between the two peaks, such that it simply provides an overall scale for the spectrum. In contrast, any change in $T$ will affect only the mechanical resonance at $\Omega_\text{m}$ through $S_{xx}(\omega)$. If the phase modulation conditions are held constant, we can then find the temperature of the mode through the ratio

\begin{equation}
R_{T}=\frac{A_{\text{m}}}{A_{\text{mod}}}\propto T
\end{equation}

\noindent where $A_\text{mod}$ is the integrated area under the phase modulation peak and $A_\text{m}$ is the area under the mechanical resonance. For the purpose of these calculations we subtract the detection noise floor, $S_{VV}^\text{NF}$.

\begin{figure}
	\includegraphics[width=\columnwidth]{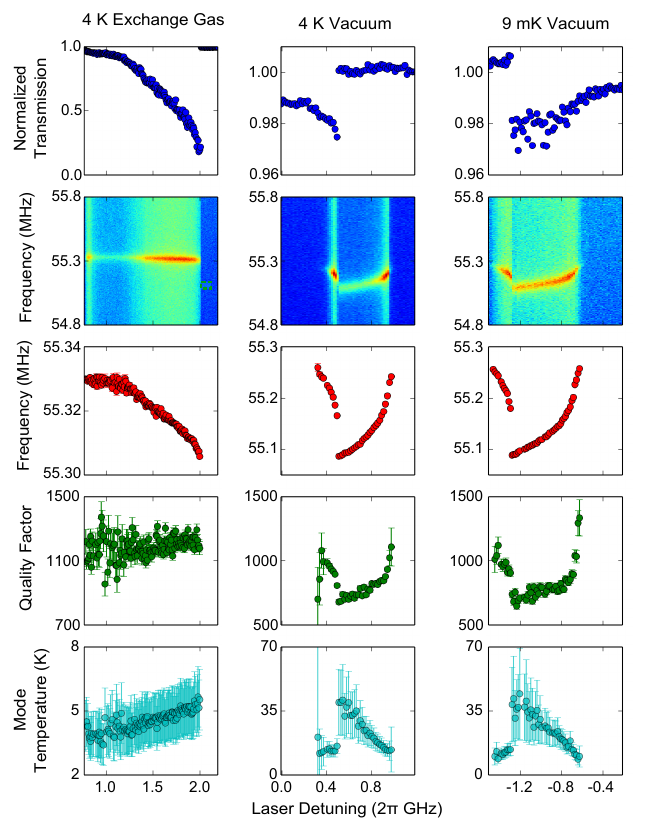}
	\caption{(From top to bottom) The low-frequency transmission of light through the tapered fiber traces out the optical resonance profile (top), while information about the mechanical resonance at 55 MHz is contained in the power spectral density of the high-frequency transmission (plotted here on a log scale). From fits to these spectra, we extract the resonance frequency (red), quality factor (green) and temperature (cyan) of the mechanical mode across the optical resonance.}
	\label{fig:AllParams}
\end{figure}

We drive our EOM at $\Omega_\text{mod}/2\pi=54.5$ MHz and measure $R_T$ at liquid helium temperature (4.2 K), where a copious quantity of exchange gas in the IVC ensures good thermalization of the bottle resonator with the outer helium bath. The result, $R_{4.2}$, fixes the temperature measurement scale. The mechanical mode temperature is thus given by

\begin{equation}
T=\frac{R_T}{R_{4.2}}\times4.2\text{ K}.
\end{equation}

Measurements of the bottle resonator made under various experimental conditions at low temperatures are shown in Figure \ref{fig:AllParams}. The laser was scanned across the optical resonance and the low- and high-frequency parts of the taper transmission were recorded simultaneously. From left to right, measurements were made at 4.2 K in exchange gas and in vacuum, as well as in vacuum with the fridge operating at its base temperature of 9 mK. In all cases, an injected power of $P_{\text{in}}=24$ $\mu$W was used. 

As the laser is tuned to the center of the optical resonance, the low-frequency transmission decreases and more photons are coupled into the bottle. There is a corresponding decrease in the mechanical resonance frequency, which is accompanied by an increase in the mode temperature. In exchange gas, the temperature increase is small, on the order of a few kelvin, and the maximum relative frequency shift amounts to approximately $-0.04\%$. In contrast, when the IVC is evacuated, we see strikingly similar behavior regardless of whether the fridge is operated at 4.2 K or base temperature. In both cases, the temperature increases to approximately 40 K, while the mechanical frequency decreases by $0.3\%$. We also observe a significant decrease in $Q_\text{m}$, from 1100 to 600, as the laser is tuned to the center of the optical resonance.

\begin{figure}
	\includegraphics[width=\columnwidth]{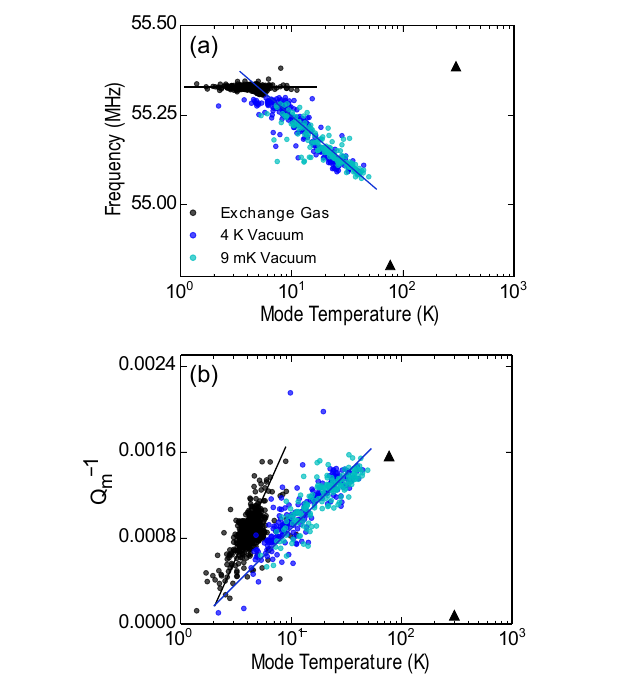}
	\caption{(a) Frequency and (b) inverse $Q_\text{m}$ versus calibrated mode temperature. Lines are guides to the eye and black triangles are data points at 77 K (liquid nitrogen) and 295 K (room temperature) in nitrogen exchange gas.}
	\label{fig:AllData}
\end{figure}

For a more direct comparison, the mechanical resonance frequency and inverse $Q_\text{m}$ are plotted versus measured mechanical mode temperature in Figure \ref{fig:AllData} for a number of injected optical powers. Black triangles indicate measurements taken at 77 K (liquid nitrogen) and 295 K (room temperature), where the bottles were thermalized using nitrogen exchange gas.

\section{Discussion}

Our measurements reveal key information about heat dissipation in the bottle resonators. Despite efforts to thermally anchor the bottles to the base temperature of the fridge, the lowest achieved mode temperature was approximately 4 K, corresponding to an average occupation of just $\sim$1500 phonons. This temperature was reached when helium exchange gas was added to the IVC, indicating that the gaseous helium facilitated convection between the resonator and the bath of liquid helium. In vacuum, regardless of whether the fridge was operated at base temperature or liquid helium temperature, the temperature of the bottle was raised upon optomechanical measurements, as shown in Figures \ref{fig:AllParams} and \ref{fig:AllData}. This is evidence of the intrinsically weak thermal connection between the mixing chamber and the silica bottle resonator.

This heating arises as a result of the absorption of laser light into the bulk silica. The degree of nonlinearity in the optical resonances, quantified through the parameter $\Delta_\text{nl}$, serves as a measurement of the temperature change generated by the absorbed light. We find that $\Delta_\text{nl}$ is nearly an order of magnitude larger in vacuum than in exchange gas, since the lack of convection greatly reduces the efficiency with which the bottle can dissipate heat. These results are entirely consistent with measurements of the mechanical frequency shift with temperature. As the laser is scanned across an optical resonance, we find that the mechanical resonance frequency exhibits the same dependence on laser detuning as the taper transmission. We furthermore observe a relative frequency shift in vacuum that is nearly ten times larger than that in exchange gas. Since the optical and mechanical modes occupy nearly the same volume within the bottle structure, heating of the optical modes by laser absorption is reflected equally in the mechanical mode.

Finally, we observe a significant dependence of the mechanical quality factor on temperature, with a shape that is characteristic of phonon-coupling to configurational two-level systems in glass \cite{Classen1994}. At high temperatures, thermal activation over a potential barrier allows transitions between the two configurations; as the temperature is lowered, thermal activation ceases and quantum tunneling can occur. At still lower temperatures, quantum tunneling is also forbidden and the mechanical quality factor increases dramatically. From the shape of Figure \ref{fig:AllData}(b), we deduce that our system resides in the region of thermal activation. Lower temperatures would thus allow us to increase the mechanical quality factor far beyond its room temperature value. There is also an increase in the mechanical damping rate when the IVC is filled with exchange gas; we attribute this to interactions with a thin film of liquid helium on the surface of the bottle. This may be in fact be a promising tool to study ultra-thin superfluid films \cite{Xu1990} via interactions with the optical and mechanical modes of resonator. We note that the general shapes of both Figure \ref{fig:AllData}(a) and (b) are in agreement with the observations of silica toroid resonators reported in Ref. \citenum{Arcizet2009}.
 
\section{Conclusion}

We have demonstrated passive cooling of a 64 ng optomechanical resonator down to just $\sim$1500 phonons. Further cooling is prevented by the inability to dissipate the heat caused by optical absorption, exacerbated by the high $Q_\text{o}$ of the silica resonator. The scale of this optical absorption was found to be in excellent agreement with the degree of heating in the mechanical mode. This comparison was enabled by optomechanical mode thermometry, detailed in the Appendix, which is now an important tool for quantum optomechanics. Future experiments will focus on improving the coupling of these sideband-resolved optomechanical resonators to the thermal bath. One possibility to achieve this is to use a local reservoir of helium as a heat link to the 9 mK dilution refrigerator environment, which would reduce optically-induced heating of the resonator \cite{Treussart1998}. In particular, it would be intriguing to use liquid helium, which has been shown to provide an excellent thermalization medium for micro-electromechanical systems (MEMS) down to $\sim$60 mK \cite{Gonzalez2013}. Furthermore, the MEMS in Ref.~\citenum{Gonzalez2013} regain their vacuum mechanical dissipation levels at $T\lesssim$100 mK, due to the temperature-dependent phonon occupation in the superfluid state. Successful thermalization of the presented nanogram-scale microresonators to 9 mK would result in average phonon occupancies of $\bar{n}\lesssim3$, while maintaining the high mechanical $Q$ and sideband-resolved nature of these optomechanical devices, opening up the door for ground state cooling \cite{Chan2011} and further quantum optomechanical protocols \cite{Palomaki2013}.  Finally, extension of these cryogenic silica microresonators to doped optical glasses would also enable new quantum technologies, such as photonic memories for quantum cryptographic networks \cite{Saglamyurek2015}.

\section{Acknowledgements}

This work was supported by the University of Alberta, Faculty of Science; Alberta Innovates Technology Futures; the Natural Sciences and Engineering Research Council, Canada; the Canada Foundation for Innovation; and the Alfred P. Sloan Foundation.

\section{Appendix: Mechanical Mode Thermometry}

\subsection{Solution of the Optomechanical Cavity}

The equation of motion for the intracavity optical field $a(t)$ in an optical cavity that is coupled to a mechanical resonator with strength $G=d\omega_\text{o}/dx$ is given by \cite{Aspelmeyer2014}

\begin{equation}
\dot{a}(t) = -\left(\frac{\kappa}{2}-i\Delta-iGx(t)\right)a(t)+\sqrt{\kappa_{\text{ex}}}s_{\text{in}}(t),
\label{eq:eom_cavity}
\end{equation}

\noindent in a frame rotating at the laser frequency $\omega_\text{L}$. Here, $s_{\text{in}}(t)$ is the input optical field, normalized such that $|s_{\text{in}}(t)|^2=\dot{n}_\text{in}=P_{\text{in}}/\hbar\omega_{\text{L}}$. The field output by the optical cavity is then

\begin{equation}
s_{\text{out}}(t)=s_{\text{in}}(t)-\sqrt{\kappa_{\text{ex}}}a(t).
\label{eq:sout}
\end{equation}

This system can be solved in the stationary regime (see for example, Ref. \citenum{Aspelmeyer2014}) but since we intend to inject a time-varying input field, we solve Equation \eqref{eq:eom_cavity} without assuming a stationary state. As in Ref.~\citenum{Schliesser2008}, we assume that $a(t)$ can be written as the sum of the solution $a_{\text{h}}(t)$ to the associated homogeneous problem and a particular solution $a_{\text{p}}(t)$,

\begin{equation}
a(t)=a_{\text{h}}(t)+a_{\text{p}}(t).
\label{eq:a_sum}
\end{equation}

\noindent We further make the assumption that the particular solution can be written as $a_{\text{p}}(t)=a_{\text{h}}(t)f(t)$ where $f(t)$ is a yet-to-be-determined function of time.

We begin by solving for $a_{\text{h}}(t)$ by taking $s_{\text{in}}(t)=0$. The resulting differential equation is

\begin{equation}
\dot{a}_\text{h}(t)=-\left(\frac{\kappa}{2}-i\Delta-iGx(t)\right)a_\text{h}(t),
\label{eq:ah_diff}
\end{equation}

\noindent with the solution

\begin{equation}
a_{\text{h}}(t)=a_0\exp\left[-\left(\frac{\kappa}{2}-i\Delta\right)t+iG\int x(t)dt\right],
\label{eq:ah_1}
\end{equation}

\noindent where $a_0$ is an amplitude set by the initial conditions of the problem. We choose the form 

\begin{equation}
x(t)=x_0e^{-\Gamma t/2}\cos\Omega_{\text{m}}t
\end{equation}

\noindent for a damped harmonic oscillator with angular frequency $\Omega_{\text{m}}$, damping rate $\Gamma$, effective mass $m_{\text{eff}}$, and peak amplitude $x_0=\sqrt{2k_{\text{B}}T/m_{\text{eff}}\Omega_{\text{m}}^2}$ at temperature $T$. Integration yields

\begin{equation}
\begin{split}
\int x(t) dt &= \frac{x_0e^{-\Gamma t/2}}{\frac{\Gamma^2}{4}+\Omega_{\text{m}}^2}\left(\Omega_{\text{m}}\sin\Omega_{\text{m}}t-\frac{\Gamma}{2}\cos\Omega_{\text{m}}t\right) \\
&\approx \frac{x_0}{\Omega_{\text{m}}}e^{-\Gamma t/2}\sin\Omega_{\text{m}}t,
\end{split}
\label{eq:x_int}
\end{equation}

\noindent where we have used the high-$Q$ approximation ($\Omega_{\text{m}}\gg\Gamma$) to neglect the cosine term.

Substituting Equation \eqref{eq:x_int} into Equation \eqref{eq:ah_1}, we have 

\begin{equation}
a_{\text{h}}(t) \approx a_0e^{-\left(\frac{\kappa}{2}-i\Delta \right)t}\left(1+\frac{\xi}{2}e^{-\Gamma t/2}\left[e^{i\Omega_{\text{m}}t}-e^{-i\Omega_{\text{m}}t}\right]\right),
\label{eq:ah_soln}
\end{equation}

\noindent where we have written 

\begin{equation}
\xi\equiv\frac{Gx_0}{\Omega_{\text{m}}}
\label{eq:xi}
\end{equation}

\noindent and used the Jacobi-Anger expression to write

\begin{equation}
e^{i\Lambda\sin\phi}=\sum\limits_{k=-\infty}^{+\infty}J_k(\Lambda)e^{ik\phi}\approx 1+i\Lambda\sin\phi
\label{eq:approx}
\end{equation}

\noindent for small $\Lambda=\xi e^{-\Gamma t/2}\ll1$. Here, $J_k(\Lambda)$ are the $k$th Bessel functions of the first kind. 

We justify this approximation by first noting that $e^{-\Gamma t/2}<1$ for all finite positive times and then rewriting $\xi$ as 

\begin{equation}
\xi=\frac{2g_0\sqrt{\frac{k_{\text{B}}T}{\hbar\Omega_{\text{m}}}}}{\Omega_{\text{m}}}\approx\frac{2g_0\sqrt{\bar{n}}}{\Omega_{\text{m}}}
\end{equation}

\noindent using the vacuum optomechanical coupling rate $g_0=Gx_{\text{zpf}}$ and the amplitude of the mechanical zero-point fluctuations $x_{\text{zpf}}=\sqrt{\hbar/2m_{\text{eff}}\Omega_{\text{m}}}$. Here, $\bar{n}=k_\text{B}T/\hbar\Omega_\text{m}$ is the average phonon occupation of the resonator for $k_{\text{B}}T\gg\hbar\Omega_{\text{m}}$. In this form, $g_0\ll\Omega_\text{m}$ is absolutely necessary to have $\xi\ll1$, and we must additionally consider the phonon occupation $\bar{n}$. For optical frequency optomechanical devices, this condition is commonly satisfied, especially at low temperatures where $\bar{n}$ is small. For the bottle resonators used in our experiments, $g_0/2\pi\sim350$ Hz, so $\xi\ll1$ and thus $\xi e^{-\Gamma t/2}\ll1$ as long as $\bar{n}\ll10^{10}$ (corresponding to a temperature of $10^{7}$ K).

We now return to Equation \eqref{eq:a_sum} to look for the particular solution to Equation \eqref{eq:eom_cavity}. We note that

\begin{equation}
\begin{split}
\dot{a}(t)&=\dot{a}_{\text{h}}(t)+\dot{a}_{\text{h}}(t)f(t)+a_{\text{h}}(t)\dot{f}(t) \\
&=-\left(\frac{\kappa}{2}-i\Delta-iGx(t)\right)\left(a_{\text{h}}(t)+a_{\text{h}}(t)f(t)\right)+\sqrt{\kappa_{\text{ex}}}s_{\text{in}}(t).
\end{split}
\end{equation}

\noindent Given that the homogeneous solution obeys Equation \eqref{eq:ah_diff}, it follows that 

\begin{equation}
\dot{f}(t) = \frac{\sqrt{\kappa_{\text{ex}}}s_{\text{in}}(t)}{a_{\text{h}}(t)}.
\label{eq:fdot}
\end{equation}

For the mechanical mode thermometry, we phase modulate the input laser light by driving an electro-optic modulator with a sinusoidal signal of the form $V_0e^{-\gamma t /2}\sin\Omega_{\text{mod}}t$, where $V_0$ is the drive voltage amplitude, $\Omega_{\text{mod}}$ is the drive frequency and $\gamma\ll\Gamma$ is the linewidth of the driving source. This produces an input field of the form 

\begin{equation}
\begin{split}
s_\text{in}(t) &= s_\text{in}e^{i\beta e^{-\gamma t/2}\sin\Omega_\text{mod}t} \\
&=s_\text{in}\sum\limits_{k=-\infty}^{+\infty}J_k(\beta e^{-\gamma t/2})e^{ik\Omega_\text{mod}t}
\end{split}
\end{equation}

\noindent where we define the phase modulation depth as $\beta=\pi V_0/V_\pi$, given the device-dependent half-wave voltage, $V_\pi$. If the phase modulation is weak ($\beta\ll1$), it is sufficient to consider only the first order sidebands at $\pm\Omega_\text{mod}$. In this case, we can again use the approximation in Equation \eqref{eq:approx} to write

\begin{equation}
s_\text{in}(t)=s_\text{in}\left(1+\frac{\beta}{2}e^{-\gamma t/2}\left[e^{i\Omega_\text{mod}t}-e^{-i\Omega_\text{mod}t}\right]\right).
\label{eq:sin}
\end{equation}

We substitute this result, along with the homogeneous solution of Equation \eqref{eq:ah_soln}, into Equation \eqref{eq:fdot}, yielding

\begin{equation}
\begin{split}
\dot{f}(t) =& \frac{\sqrt{\kappa_\text{ex}}s_\text{in}}{a_0}e^{\left(\frac{\kappa}{2}-i\Delta\right)t}\left(1+\frac{\beta}{2}e^{-\gamma t/2}\left[e^{i\Omega_\text{mod}t}-e^{-i\Omega_\text{mod}t}\right]\right) \\
&\times\left(1-\frac{\xi}{2}e^{-\Gamma t/2}\left[e^{i\Omega_\text{m}t}-e^{-i\Omega_\text{m}t}\right]\right).
\end{split}
\label{eq:fdot_1}
\end{equation}

\noindent Keeping only terms to first order in the small parameters $\xi$ and $\beta$, integrating Equation \eqref{eq:fdot_1} and multiplying by $a_\text{h}(t)$ (Equation \eqref{eq:ah_soln}), we obtain the particular solution

\begin{equation}
\begin{split}
a_\text{p}(t)=& \sqrt{\kappa_{\text{ex}}}s_{\text{in}}\left(\frac{1}{\frac{\kappa}{2}-i\Delta}+\frac{\beta}{2}e^{-\frac{\gamma t}{2}}\left[\frac{e^{i\Omega_{\text{mod}}t}}{\frac{\kappa}{2}-i(\Delta-\Omega_{\text{mod}})} \right.\right. \\ &\left.\left.-\frac{e^{-i\Omega_{\text{mod}}t}}{\frac{\kappa}{2}-i(\Delta+\Omega_{\text{mod}})}\right]-\frac{\xi}{2}e^{-\frac{\Gamma t}{2}}\left[\frac{e^{i\Omega_{\text{m}}t}}{\frac{\kappa}{2}-i(\Delta-\Omega_{\text{m}})} \right.\right.\\
&\left.\left.-\frac{e^{-i\Omega_{\text{m}}t}}{\frac{\kappa}{2}-i(\Delta+\Omega_{\text{m}})}-\frac{e^{i\Omega_{\text{m}}t}}{\frac{\kappa}{2}-i\Delta}+\frac{e^{-i\Omega_{\text{m}}t}}{\frac{\kappa}{2}-i\Delta}\right]\right),
\end{split}
\label{eq:ap_soln}
\end{equation}

\noindent where we have used the fact that $\kappa\gg\Gamma,\gamma$ to simplify the final expression. We note that $a_\text{h}(t)$ decays much more rapidly than $a_\text{p}(t)$, so we neglect $a_\text{h}(t)$ and take $a(t)\approx a_\text{p}(t)$. 

We now calculate the field output by the cavity by substituting Equations \eqref{eq:sin} and \eqref{eq:ap_soln} into Equation \eqref{eq:sout},

\begin{equation}
\begin{split}
s_{\text{out}}(t)=& s_{\text{in}}\left(1+\frac{\beta}{2}e^{-\gamma t/2}\left[e^{i\Omega_{\text{mod}}t}-e^{-i\Omega_{\text{mod}}t}\right]-\frac{\kappa_{\text{ex}}}{\frac{\kappa}{2}-i\Delta} \right. \\
&-\frac{\kappa_{\text{ex}}\beta}{2}e^{-\frac{\gamma t}{2}}\left[\frac{e^{i\Omega_{\text{mod}}t}}{\frac{\kappa}{2}-i(\Delta-\Omega_{\text{mod}})} -\frac{e^{-i\Omega_{\text{mod}}t}}{\frac{\kappa}{2}-i(\Delta+\Omega_{\text{mod}})}\right] \\
& \left.+\frac{\kappa_{\text{ex}}\xi}{2}e^{-\frac{\Gamma t}{2}}\left[\frac{e^{i\Omega_{\text{m}}t}}{\frac{\kappa}{2}-i(\Delta-\Omega_{\text{m}})} -\frac{e^{-i\Omega_{\text{m}}t}}{\frac{\kappa}{2}-i(\Delta+\Omega_{\text{m}})}\right.\right.\\
&\left.\left.-\frac{e^{i\Omega_{\text{m}}t}}{\frac{\kappa}{2}-i\Delta}+\frac{e^{-i\Omega_{\text{m}}t}}{\frac{\kappa}{2}-i\Delta}\right]\right).
\end{split}
\label{eq:sout_PM}
\end{equation}

\subsection{Detection}

In our experiments, we use a direct detection scheme where a photodetector sensitive to the intensity of the light outputs a voltage signal $V(t)$ proportional to $|s_\text{out}(t)|^2$, 

\begin{equation}
V(t) = V_\text{DC}+V_\text{m}(t)+V_\text{mod}(t).
\end{equation}

\noindent We have separated the output signal into a part 

\begin{equation}
V_\text{DC} = H(0)\dot{n}_\text{in}\left(1-\frac{\kappa_0\kappa_\text{ex}}{\frac{\kappa^2}{4}+\Delta^2}\right)
\end{equation}

\noindent which is constant in time, and two high-frequency parts which oscillate at the mechanical resonance frequency,

\begin{equation}
\begin{split}
V_\text{m}(t) =& H(\Omega_\text{m})\dot{n}_\text{in}\kappa_\text{ex}\frac{\xi}{2} e^{-\Gamma t/2}\left(\frac{C_1(\Omega_\text{m},\Delta)\cos\Omega_\text{m}t}{D(0,\Delta)D(\Omega_\text{m},\Delta)D(-\Omega_\text{m},\Delta)}\right.\\
&\left.+\frac{C_2(\Omega_\text{m},\Delta)\sin\Omega_\text{m}t}{D(0,\Delta)D(\Omega_\text{m},\Delta)D(-\Omega_\text{m},\Delta)}\right),
\end{split}
\label{eq:Vm_t}
\end{equation}

\noindent and at the modulation frequency, 

\begin{equation}
\begin{split}
V_\text{mod}(t) =& -H(\Omega_\text{mod})\dot{n}_\text{in}\kappa_\text{ex}\frac{\beta}{2} e^{-\gamma t/2}\times\\
&\left(\frac{C_1(\Omega_\text{mod},\Delta)\cos\Omega_\text{mod}t}{D(0,\Delta)D(\Omega_\text{mod},\Delta)D(-\Omega_\text{mod},\Delta)}\right. \\
&\left.+\frac{C_2(\Omega_\text{mod},\Delta)\sin\Omega_\text{mod}t}{D(0,\Delta)D(\Omega_\text{mod})D(-\Omega_\text{mod})}\right),
\end{split}
\label{eq:Vmod_t}
\end{equation}

\noindent respectively. $H(\omega)$ is a function which describes the frequency response of the photodetector and other detection electronics. We assume that it is locally flat, such that for a suitably chosen $\Omega_\text{mod}$, we can take $H(\Omega_\text{mod})=H(\Omega_\text{m})=H$. Note that $V_\text{DC}$ is simply the transmission profile for the optical resonance, as given by Equations \eqref{eq:nout} and \eqref{eq:n_cav} in the main text. For convenience, we have also defined 

\begin{equation}
C_1(\Omega,\Delta) = \Delta\Omega(-\kappa[\kappa^2+4\Delta^2]+\kappa_\text{ex}[\kappa^2+4\Delta^2-4\Omega^2]),
\end{equation}

\begin{equation}
C_2(\Omega,\Delta) = \Delta\Omega^2(\kappa[4\kappa_\text{ex}-3\kappa]+4[\Delta^2-\Omega^2]),
\end{equation}

\noindent and 

\begin{equation}
D(\Omega,\Delta) = \frac{\kappa^2}{4}+(\Delta-\Omega)^2.
\end{equation}

To obtain the power spectral density of the voltage signal, we must first Fourier transform $V(t)$. In our experiment, this is done as part of the software post-processing, but it can also be performed in hardware, using a network or spectrum analyzer, for example. $V_\text{DC}$ is filtered out by a high-pass filter on our photodetector so we address only the high-frequency part of $V(t)$. 

We define the Fourier transform as 

\begin{equation}
V(\omega) = \int\limits_{-\infty}^{\infty}e^{-i\omega t}V(t)dt.
\end{equation}

\noindent Since the Fourier transform is linear, we apply it to $V_\text{m}(t)$ and $V_\text{mod}(t)$ separately. We begin by substituting the form of $\xi$ given by Equation \eqref{eq:xi} into Equation \eqref{eq:Vm_t} and noting that 

\begin{equation}
\begin{split}
\dot{x}(t) &= -\frac{\Gamma x_0}{2}e^{-\Gamma t/2}\cos\Omega_\text{m}t-\Omega_\text{m}x_0e^{-\Gamma t/2}\sin\Omega_\text{m}t \\
&\approx-\Omega_\text{m}x_0e^{-\Gamma t/2}\sin\Omega_\text{m}t.
\end{split}
\end{equation}

\noindent Equation \eqref{eq:Vm_t} then becomes

\begin{equation}
V_\text{m}(t) = H\dot{n}_\text{in}\frac{G\kappa_\text{ex}}{2\Omega_\text{m}}\left(\frac{C_1(\Omega_\text{m},\Delta)x(t)-C_2(\Omega_\text{m},\Delta)\dot{x}(t)/\Omega_\text{m}}{D(0,\Delta)D(\Omega_\text{m},\Delta)D(-\Omega_\text{m},\Delta)}\right),
\end{equation}

\noindent with its Fourier transform given by 

\begin{equation}
V_\text{m}(\omega) = H\dot{n}_\text{in}\frac{G\kappa_\text{ex}}{2\Omega_\text{m}}\left(\frac{C_1(\Omega_\text{m},\Delta)-i\omega C_2(\Omega_\text{m},\Delta)/\Omega_\text{m}}{D(0,\Delta)D(\Omega_\text{m},\Delta)D(-\Omega_\text{m},\Delta)}\right)X(\omega).
\end{equation}

\noindent Near the mechanical resonance frequency, $\omega\approx\Omega_\text{m}$ and 

\begin{equation}
V_\text{m}(\omega) = H\dot{n}_\text{in}\frac{G\kappa_\text{ex}}{2\Omega_\text{m}}\left(\frac{C_1(\Omega_\text{m},\Delta)-i C_2(\Omega_\text{m},\Delta)}{D(0,\Delta)D(\Omega_\text{m},\Delta)D(-\Omega_\text{m},\Delta)}\right)X(\omega).
\label{eq:Vm_omega}
\end{equation}

Similarly, $V_\text{mod}(t)$ can be rewritten in terms of the original phase modulation signal $\phi(t) = \beta e^{-\gamma t/2}\sin\Omega_\text{mod}t$, 

\begin{equation}
\begin{split}
V_\text{mod}(t) =& -H\dot{n}_\text{in}\frac{\kappa_\text{ex}}{2}\left(\frac{C_1(\Omega_\text{mod},\Delta)\dot{\phi}(t)/\Omega_\text{mod}}{D(0,\Delta)D(\Omega_\text{mod},\Delta)D(-\Omega_\text{mod},\Delta)}\right.\\
	&\left.+\frac{C_2(\Omega_\text{mod},\Delta)\phi(t)}{D(0,\Delta)D(\Omega_\text{mod},\Delta)D(-\Omega_\text{mod},\Delta)}\right).
\end{split}
\label{eq:Vmod_t2}
\end{equation}

\noindent Analogous to Equation \eqref{eq:Vm_omega}, the Fourier transform of Equation \eqref{eq:Vmod_t2} near the modulation frequency is 

\begin{equation}
\begin{split}
V_\text{mod}(\omega) =& -H\dot{n}_\text{in}\frac{\kappa_\text{ex}}{2}\left(\frac{iC_1(\Omega_\text{mod},\Delta)}{D(0,\Delta)D(\Omega_\text{mod},\Delta)D(-\Omega_\text{mod},\Delta)}\right.\\
&\left.+ \frac{C_2(\Omega_\text{mod},\Delta)}{D(0,\Delta)D(\Omega_\text{mod},\Delta)D(-\Omega_\text{mod},\Delta)}\right)\Phi(\omega),
\end{split}
\label{eq:Vmod_omega}
\end{equation}

\noindent where $\Phi(\omega)$ is the Fourier transform of $\phi(t)$. 

\begin{figure}[t]
	\includegraphics[width=\columnwidth]{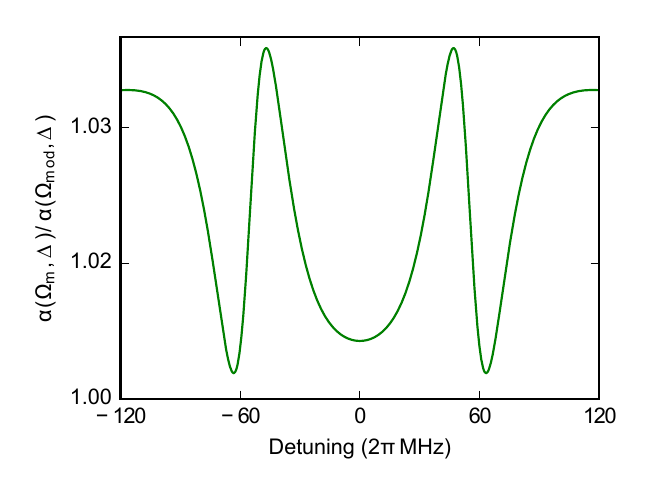}
	\caption{Calculated ratio of $\alpha(\Omega_\text{m},\Delta)$ to $\alpha(\Omega_\text{mod},\Delta)$ for experimental parameters ($\kappa/2\pi=50$ MHz, $\kappa_\text{ex}/2\pi=20$ MHz, $\Omega_\text{m}/2\pi=55$ MHz and $\Omega_\text{mod}/2\pi=54.5$ MHz). A variation of less than 3\% in this ratio across the optical resonance is shown.}
	\label{fig:ratio}
\end{figure}

Combining the results of Equations \eqref{eq:Vm_omega} and \eqref{eq:Vmod_omega}, we calculate the voltage power spectral density,

\begin{equation}
S_{VV}(\omega) = \lim\limits_{\tau\to\infty}\frac{1}{\tau}|V_\text{m}(\omega)+V_\text{mod}(\omega)|^2,
\end{equation}

\noindent which can be reduced to 

\begin{equation}
S_{VV}(\omega) = \lim\limits_{\tau\to\infty}\frac{1}{\tau}|V_\text{m}(\omega)|^2+\lim\limits_{\tau\to\infty}\frac{1}{\tau}|V_\text{mod}(\omega)|^2
\end{equation}

\noindent if we choose $\Omega_\text{mod}$ such that $|\Omega_\text{mod}-\Omega_\text{m}|\gg\Gamma,\gamma$. This prevents overlap between the peaks in the spectrum and allows us to neglect the cross terms. The final result is then

\begin{equation}
S_{VV}(\omega) = \dot{n}_\text{in}^2\left(\frac{G^2}{\Omega_\text{m}^2}\alpha(\Omega_\text{m},\Delta)S_{xx}(\omega)+\alpha(\Omega_\text{mod},\Delta)S_{\phi\phi}(\omega)\right)
\end{equation}

\noindent where $S_{\phi\phi}(\omega)$ is the PSD of the phase fluctuations induced by the phase calibration signal. We have defined the transduction coefficient as

\begin{equation}
\alpha(\Omega,\Delta)\equiv\frac{H^2\kappa_\text{ex}^2}{4}\frac{C_1^2(\Omega,\Delta)+C_2^2(\Omega,\Delta)}{D^2(0,\Delta)D^2(\Omega,\Delta)D^2(-\Omega,\Delta)}.
\end{equation}

We see that for a choice of the modulation frequency $\Omega_\text{mod}\approx\Omega_\text{m}$, $\alpha(\Omega_\text{mod},\Delta)\approx\alpha(\Omega_\text{m},\Delta)$ and the phase modulation signal is transduced nearly identically to the mechanical motion by the optical cavity and photodetector. In actuality, $\alpha$ is also dependent on the detuning of the laser, but it can be shown that for $|\Omega_\text{m}-\Omega_\text{mod}|/\Omega_\text{m}\sim1\%$, the variation of the ratio of $\alpha(\Omega_\text{m},\Delta)/\alpha(\Omega_\text{mod},\Delta)$ with detuning is negligibly small. For the parameters in our experiment, Figure \ref{fig:ratio} illustrates that this variation across laser detuning is less than 3\% in our experiment.

Although we have focussed on direct detection techniques, this method of calculating the detected PSD applies equally well to other detection schemes, including optical homodyne or heterodyne systems. It can be shown that analogous results, albeit with a different functional dependence on the laser detuning, can be obtained for such systems.

We finally note that we have made no assumptions about the optomechanical system, beyond requiring that $\kappa\gg\Gamma$, as is true for any system in the standard hierarchy of optomechanics. The only constraints placed on the modulation signal are that it is small ($\beta\ll1$) and that its frequency is chosen appropriately, namely that $\Omega_\text{mod}$ is close enough to $\Omega_\text{m}$ that any frequency dependence in the detection electronics can be neglected and far enough that overlap between the two peaks can be ignored. It is additionally beneficial if $\Omega_\text{mod}$ is chosen close enough to $\Omega_\text{m}$ that any detuning-dependence in the ratio $\alpha(\Omega_\text{m},\Delta)/\alpha(\Omega_\text{mod},\Delta)$ can be neglected (as is true in our experiment); however, even if this is not the case, the closed form of $\alpha(\Omega,\Delta)$ allows the for the analytical calculation of this ratio provided that the optical resonance is well-characterized and the laser detuning is known.

\end{document}